\documentclass[prl,twocolumn,showpacs,amsmath,amssymb,superscriptaddress]{revtex4-1}
\usepackage{amsmath,amssymb,amsfonts,bm,color,graphicx,tabularx,physics}
\usepackage[unicode=true,colorlinks=true]{hyperref}
\usepackage[etex=true,export]{adjustbox}
\hypersetup{linkcolor=blue,citecolor=blue,urlcolor=blue}
\usepackage{tabularx,multirow,array,diagbox}
\usepackage{adjustbox}

\newcommand{\beq}{\begin{equation}}
	\newcommand{\eeq}{\end{equation}}
\newcommand{\beql}{\begin{equation*}}
	\newcommand{\eeql}{\end{equation*}}
\newcommand{\beqn}{\begin{eqnarray}}
	\newcommand{\eeqn}{\end{eqnarray}}

\begin{document}
\title{0-$\pi$ qubit in one Josephson junction}
	\author{Guo-Liang Guo}
	\author{Han-Bing Leng}
	\affiliation{School of Physics and Institute for Quantum Science and Engineering, Huazhong University of Science and Technology, Wuhan, Hubei 430074, China}
	\affiliation{Wuhan National High Magnetic Field Center and Hubei Key Laboratory of Gravitation and Quantum Physics, Wuhan, Hubei 430074, China}
	\author{Yong Hu}
		\affiliation{School of Physics and Institute for Quantum Science and Engineering, Huazhong University of Science and Technology, Wuhan, Hubei 430074, China}
	\author{Xin Liu}
	\email{phyliuxin@hust.edu.cn}
	\affiliation{School of Physics and Institute for Quantum Science and Engineering, Huazhong University of Science and Technology, Wuhan, Hubei 430074, China}
	\affiliation{Wuhan National High Magnetic Field Center and Hubei Key Laboratory of Gravitation and Quantum Physics, Wuhan, Hubei 430074, China}
\date{\today}

\begin{abstract}
Quantum states are usually fragile which makes quantum computation being not as stable as classical computation. Quantum correction codes can protect quantum states but need a large number of physical qubits to code a single logic qubit. Alternatively, the protection at the hardware level has been recently developed to maintain the coherence of the quantum information by using symmetries. However, it generally has to pay the expense of increasing the complexity of the quantum devices. In this work, we show that the protection at the hardware level can be approached without increasing the complexity of the devices. The interplay between the spin-orbit coupling and the Zeeman splitting in the semiconductor allows us to tune the Josephson coupling in terms of the spin degree of freedom of Cooper pairs, the hallmark of the superconducting spintronics. This leads to the implementation of the parity-protected 0-$\pi$ superconducting qubit with only one highly transparent superconductor-semiconductor Josephson junction, which makes our proposal immune from the various fabrication imperfections.
\end{abstract}

\maketitle

Superconducting circuits provide a promising platform for quantum computing. They utilize the Josephson effect, the coherent tunneling of Cooper pairs, to obtain the necessary anharmonicity to form superconducting qubits. Therefore, the Josephson junction is the core unit of the superconducting quantum computation. At present, the superconducting qubit based on transmon\cite{Douifmmodemboxccelseccfiot2002} has achieved high fidelity in both single-qubit and two-qubit gates \cite{Paik2011,Barends2013,Place2021,Casparis2016}. However, since its junction is composed of an insulator, the manipulation of the Josephson junction is limited to only one degree of freedom after fabrication, the Josephson coupling energy. This makes it impossible to balance the contradictory requirements for simultaneously enhancing the anharmonicity and reducing the charge noise. For transmon-like qubit, a small anharmonicity is an inevitable compromise to suppress the charge noise, although it will result in unwanted excitation to high-level states. The recently developed transmon-based 0-$\pi$ qubit has a great potential to solve this contradiction while protecting the quantum coherence at the hardware level\cite{Rymarz2021,Knill2005,Kitaev,Gyenis2021,Bell2014}. As the implementation of 0-$\pi$ qubit requires an additional controllable degree of freedom\cite{Smith2020,Dempster2014,Brooks2013,Gyenis2021,Paolo2018,Kalashnikov2020}, for the transmon-like qubit, the price paid is having to increase the complexity of the circuit. On the other hand, the Josephson effect has achieved tremendous progress over the past two decades, which is mainly due to the replacement of junction materials with semiconductors (Sms), ferromagnets (FMs), topological insulators, two-dimensional materials, etc. This not only improves the tunability of the Josephson coupling energy but also enables the multi-dimensional control of the Josephson junction.\cite{Wang2019,Gladchenko2009,Kringhoej2020,Hays2018,Bargerbos2020,Lange2015,alicki2009lattice} The former has given birth to gatemon qubits\cite{Larsen2015}, which is based on the same principle as transmon but with fully electrical control, and is leading to more transmon variants. The latter not only has spawned the field of superconducting spintronics but also is benefiting many other fields such as topological quantum computing.

\begin{figure}[t]
\centerline{\includegraphics[width=1\columnwidth]{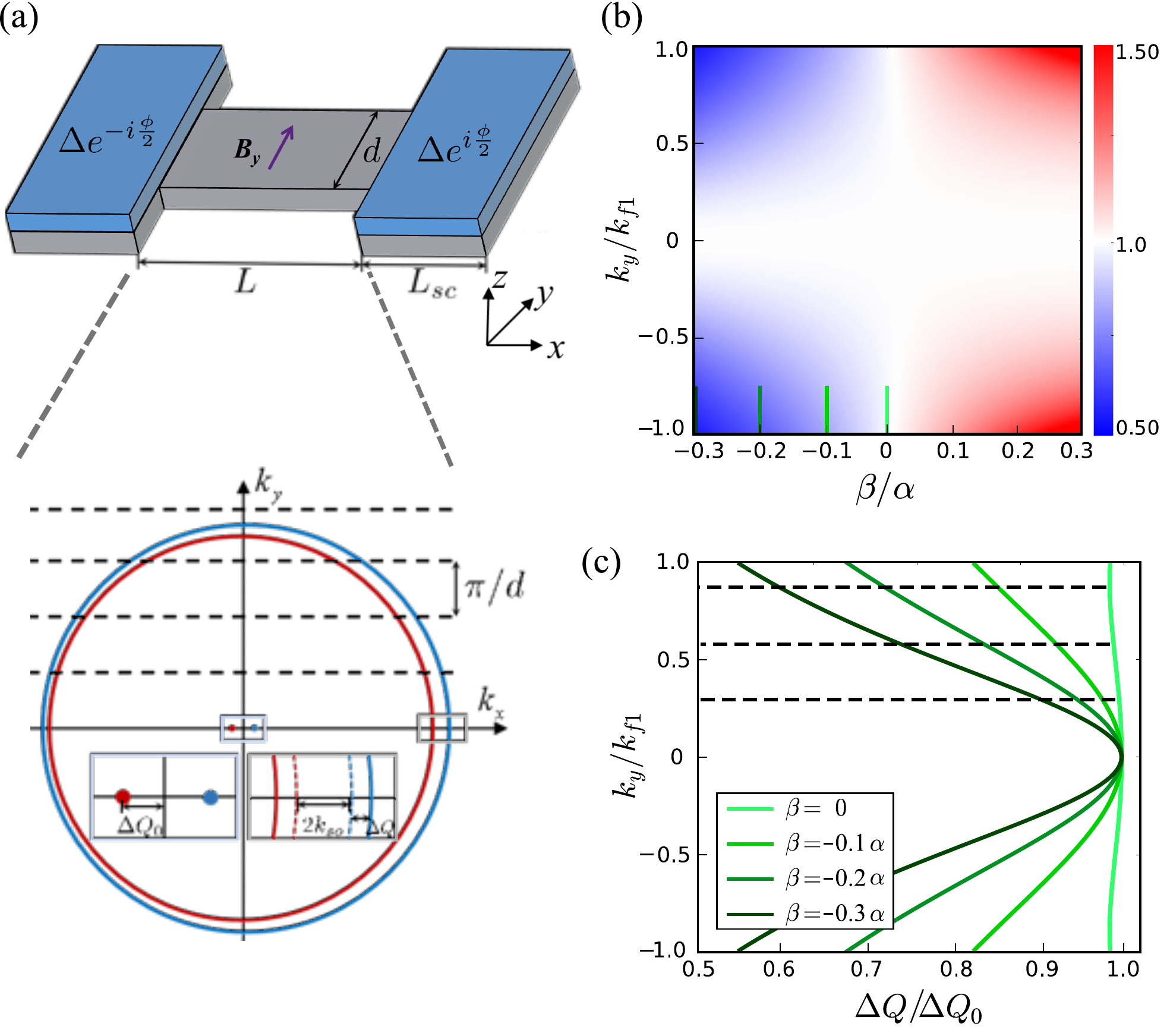}}
\caption{(a) The setup of SC/Sm/SC Josephson junction with a few transverse channels. The red and blue curves indicate the Fermi surfaces with opposite chirality due to the Rashba SOC. The solid (dashed) curves in the inset show the Fermi surfaces with (without) in-plane magnetic field. The in-plane magnetic field shifts the Fermi surfaces with opposite chirality to the opposite direction. $\Delta Q_0$ is the shift for $k_y=0$ channel. The shift of the Fermi surfaces (normalized by $\Delta Q_0$) is plot in (b) as a function of both the momentum $k_y$ and the ratio, $\beta/\alpha$, between the Dresselhaus and Rashba SOC strengths, and in (c) along the line cuts with $\beta/\alpha=0, 0.1, 0.2, 0.3$. }
 \label{set_up}
\end{figure}

In this work, we propose to implement a 0-$\pi$ qubit in only one Josephson junction (Fig.~\ref{set_up}(a)), utilizing the spin degrees of freedom inside the junction. The spin splitting in the semiconductor region provide two Fermi surfaces which can be taken as two effective Josephson junctions with almost identical Josephson coupling energy. This identity is robust against various fabrication and control imperfections such as gate voltage fluctuations, disorders, etc. The interplay, among the spin-orbit coupling, Zeeman effect, and superconductivity, induces the spin-singlet and spin-triplet Cooper pairs transition through the quantum interference between two Fermi surfaces. The width of the junction is required to be very narrow with only a few transverse modes so that this interplay can suppress the single Cooper pair tunneling and realize the degenerated $0-\pi$ qubit states. Gatemon-like qubit is shown to be an ideal platform to realize this proposal. Finally, with the practical experimental parameters, we show that the qubit relaxation time $T_1$ and coherent time $T_2$ can be dramatically increased, which exhibits the potential great advantages in superconductor-semiconductor based qubits.

%
%

{\it Model-} 
The Hamiltonian of the superconducting qubit can be generally considered as the combination of the charging energy and the Josephson potential as 
\beqn
H=4E_c(\hat{n}-n_g)^2+V_J(\hat{\phi}), 
\eeqn
with $\hat{n}$ the Cooper pair number operator, $n_g$ the offset charge, and $\hat{\phi}$ the superconducting phase operator. Generally, the Josephson potential takes the form \cite{Kringhoej2018}
\beqn
V_{J}(\hat{\phi}) = \sum_n E^{(n)}_{J\alpha} \cos(n\hat{\phi}) + E^{(n)}_{J\beta} \sin(n\hat{\phi}),
\eeqn
in which the second term indicates the finite sine terms are allowed for the system breaking both time-reversal and inversion symmetries. For the superconductor/insulator/superconductor (SC/I/SC) junction, the Josephson potential is completely dominated by the form $E^{(1)}_{J\alpha}\cos\hat{\phi}$, resulting in only one tunable parameter $E^{(1)}_{J\alpha}$. Replacing the insulator with semiconductors, the Josephson energy is known to have richer controllable forms due to strong spin-orbit coupling (SOC), which splits the Fermi surface into two in the normal region. The Hamiltonian for the normal region of the SC/Sm/SC junction takes the form
\beqn
H_{\rm sm}=\frac{\hbar^2 k^2}{2m}\sigma_0  + \bm{h}(\bm k)\cdot \bm{\sigma}  + M(x) \sigma_y ,\nonumber
\eeqn
with $m$ the effective mass, $\bm{h}$ the SOC field and $M$ the Zeeman field. 
Taking the semiconductor as InAs 2DEG, there exist two types of SOCs, namely Rashba SOC and Dresselhauss SOC, which can generally takes the form   
\beqn
\bm{h}(\bm{k}) \cdot \bm{\sigma}=(\tilde{\alpha}+\tilde{\beta})k_x\sigma_y - (\tilde{\alpha}-\tilde{\beta})k_y \sigma_x,
\eeqn                                          
with $\alpha$ and $\beta$ the Rashba and Dresselhauss SOC strengths respectively \cite{Winkler2003}. In the absence of the magnetic field and taking the Rashba SOC for example, the electrons are splitted to two Fermi surfaces at the Fermi level as
\beqn
E_{\rm f}=\frac{h^2k^2}{2m} \pm \alpha k,
\eeqn
with $\alpha$ the Rashba SOC strength and $+(-)$ corresponds to the smaller (larger) Fermi surface (Fig.~\ref{set_up}(a)). As SOC respects time-reversal symmetry, the two electronic states, which are time-reversal pairs with each other, belong to the same Fermi surface. Therefore, each Fermi surface can independently support Cooper pairs with zero center-of-mass momentum and opposite spin. With adding an external magnetic field along the appropriate direction , the two Fermi surfaces will gain the opposite center-of-mass momentum $\Delta Q(k_y)$ \cite{Pientka2017,Li2019} (Fig.~\ref{set_up}(a)). In the limit $M \ll |h(k_f)|$ and $\beta=0$, the magnitude of the center-of-momentum in almost all $k_y$ channels satisfies $ |\Delta Q| \approx  M/\hbar  v_{\rm f}$. Meanwhile, in the limit $\Delta Q \ll k_f$, or equivalently $|h(k_f)| \ll \mu$, the spins for the two states with $k_y$ and $-k_y$ in the same Fermi surface remain the anti-parallel to each other up to the first order of $M/|h(k_f)|$. When the Cooper pairs enter the normal region, they will be split into two Fermi surfaces and gain opposite center-of-mass momentum $\Delta Q = \pm M/\hbar v_{\rm f}$. Therefore, in the case of $\beta \approx 0$, the two Fermi surfaces can be effectively regarded as two interference Josephson junctions.
As $\beta$ increases, $\Delta Q$ becomes more dependent on $k_y$ (Fig.~\ref{set_up}(c)). We will show later this finite Dresselhaus SOC does not affect the $0-\pi$ qubit in our proposal

\begin{figure}[t]
\centerline{\includegraphics[width=1\columnwidth]{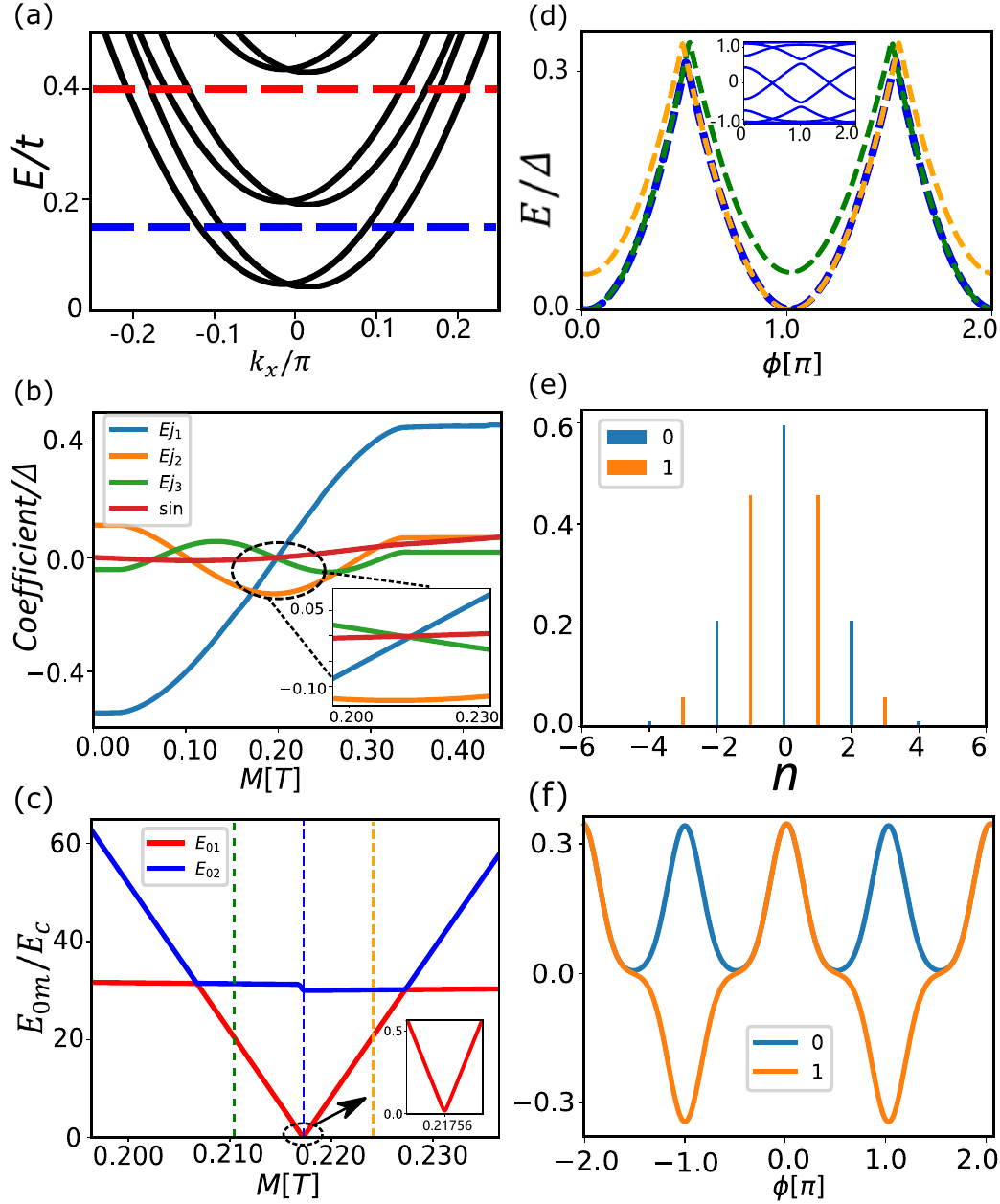}}
\caption{(a)Energy band with period condition in x-direction. (b)several leading term changes with magnetic field for the full range and inset is around 0-pi point. (c)$E_{01},E_{02}$ changes with the magnetic field around 0-pi point. (d) free energy around 0-pi point, and the three lines correspond three vertical lines in (c). (e)(f) charge distribution and phase distribution of the two lowest states. All plot at the condition $E^{(2)}_{j\alpha}/E_c\approx50,n_g=0$.}
 \label{fig2}
\end{figure}

 In the long junction limit ($d \gg \xi$), the system can support MZMs \cite{Pientka2017,Hart2017}. Here, we consider the opposite limit with $d \ll \xi$ so that there are only a few transverse channels in the normal region (indicated by the dashed lines in Fig.~\ref{set_up}(b).
 For simplicity, we first consider the case for only one channel with the chemical potential indicated by the blue dashed lines in Fig.~\ref{fig2}(a). To take into account the high-frequency Fourier series contribution, it is convenient to write the Hamiltonian in the Cooper pair number basis \cite{Bouchiat2003,alicki2009lattice}
\beqn
H= \sum_{ij} \Big[4E_c (\delta_{ij}-n_g)^2 + \sum_n \big(E_{n} \delta_{i,j+n}+h.c\big) \Big]|i \rangle\langle j|, \nonumber
\eeqn
\begin{figure}[t]
\centerline{\includegraphics[width=1\columnwidth]{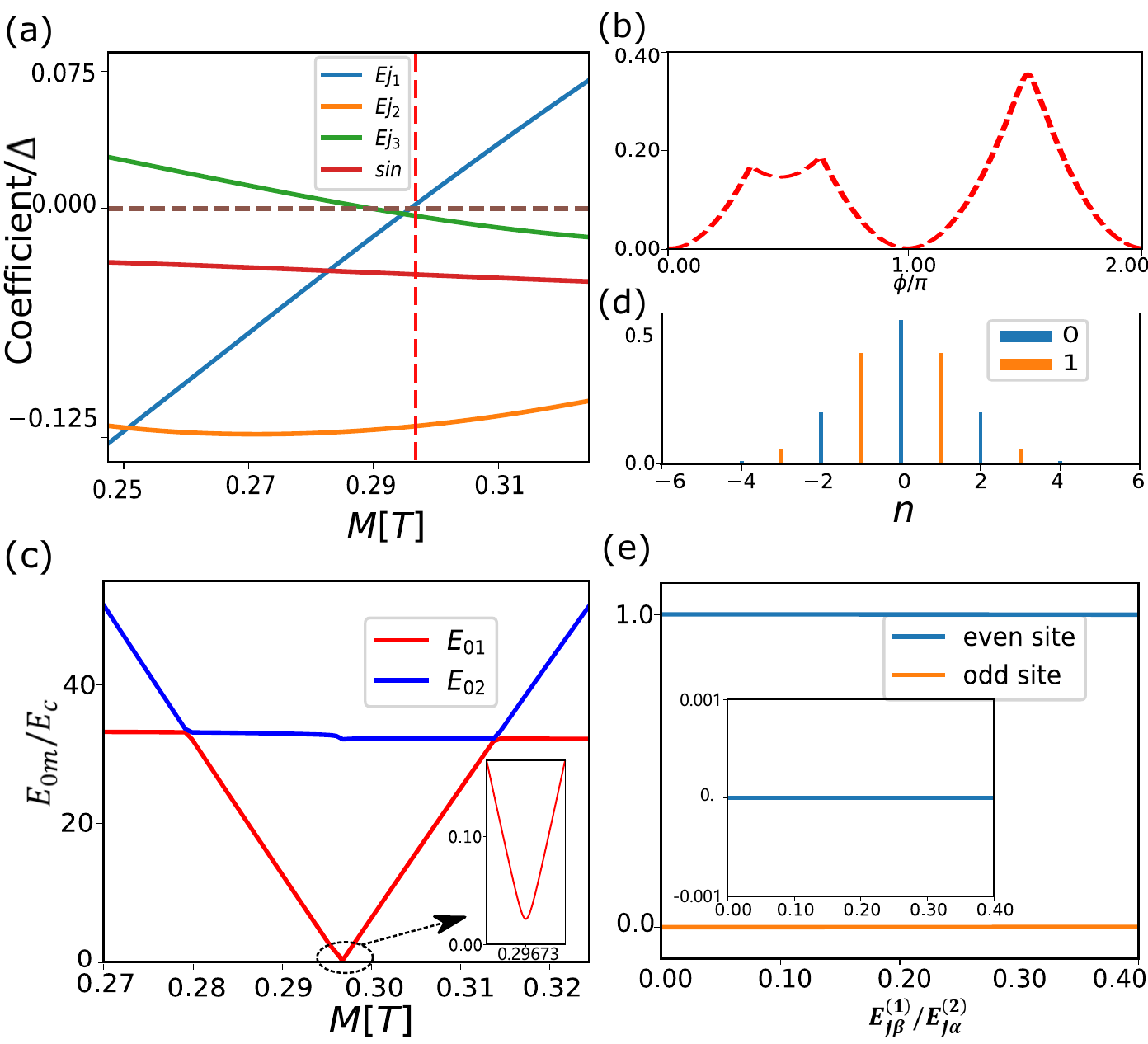}}
\caption{Two channel case and with Dresselhaus SOC $\beta=0.2\alpha$(a)several leading coefficient changes with the magnetic field around 0-pi point. (b)free energy at 0-pi point corresponds to the vertical line in (a). (c)$E_{01}, E_{02}$ changes with the magnetic field around 0-pi point (d)the charge distribution of the lowest two states. (e)the distribution of even state on even site and odd site changes with $E^{(1)}_{j\beta}/E^{(2)}_{j\alpha}$ in the condition $n_g=0,E_c=0.1GHz$, and $E^{(2)}_{j\alpha}/E_c\approx50$,inset is the matrix element of $\frac{\langle \psi_0 |E_{j\beta}^{(1)}\sin\phi| \psi_1\rangle}{E_{01}}$changes with $E^{(1)}_{j\beta}/E^{(2)}_{j\alpha}$.}
 \label{fig3}
\end{figure}
with $E_{n}=(E_{J\alpha}^{(n)}+iE_{J\beta}^{(n)})/2$ the $n$-th nearest hopping due to the $n$ Cooper pairs tunneling simultaneously. The magnitudes of the Josephson potential $E^{(n)}_{J\alpha}$ and $E^{(n)}_{J\beta}$, plotted in Fig.~\ref{fig2}(b), are obtained by calculating the Fourier components of the free energy. Note that $E^{(n)}_{J\beta}\approx 0$ in all considered magnetic field range and can be neglected. Accordingly, we show the energies of the three lowest states as $E_{01}=E_{1}-E_{0}$ and $E_{02}=E_{2}-E_0$ with finite $E_c$ and $n_g=0$ in Fig.~\ref{fig2}(c) as a function of magnetization. When the magnetization reaches the value at which the $\cos(2n+1)\phi$ terms vanishes (Fig.~\ref{fig2}(b)), the lowest two energies become almost degeneracy and has a gap with the third-lowest energy (Fig.~\ref{fig2}(c)). In this case, the Josephson junction is at the 0- to $\pi$-junction transition point, which reflects in the degenerated double-well potential of the free energy and the $\pi$-periodic Andreev levels (Fig.~\ref{fig2}(d)). Meanwhile, the Josephson coupling is dominated by the two Cooper pairs tunneling (Fig~\ref{fig2}(b)). Here and after, the minimum of $E_{01}$ refers to the lowest two eigenenergies difference at this 0- to $\pi$-junction transition point if not specifically declared. Notice that the charging energy makes the lowest two energy states not completely degenerate, but there is an energy gap proportional to $E_{c}$ (inset of Fig.~\ref{fig2}(c)). To understand these two lowest energy states at the minimum of $E_{01}$, we plot their probability density in either Cooper number basis and phase $\phi$ basis Fig.~\ref{fig2}(e)(f). It is clear that the two wave functions $|\psi_0\rangle$ and $|\psi_1\rangle$ are the eigenstates of the Cooper pair parity operator $\hat{P}=e^{i\hat{n}\pi}$ with eigenvalues $+1$ and $-1$ respectively. Meanwhile, in $\phi$ basis, each eigenstate is mainly distributed around $\phi=0$ and $\phi=\pi$ with the property
\beqn\label{wf}
|\psi_{0(1)}\rangle = \frac{|\phi \approx 0\rangle +(-) |\phi \approx \pi \rangle}{\sqrt{2}},
\eeqn
with $|\phi\approx 0\rangle$ and $|\phi\approx \pi\rangle$ refers to the state solely localized at the potential well $\phi=0$ and $\phi=\pi$ (Fig.~\ref{fig2}(b)) respectively. Therefore, we obtain a nearly degenerated qubit (Fig.~\ref{fig2}(c)), which has an energy splitting around $E_c$ and are isolated from other states by an energy gap around $\sqrt{32E^{(2)}_{j\alpha} E_c}$ \cite{Koch2007}.

Considering the experimental reality, there may be multiple channels in the normal region and finite Dresselhaus SOC, which can complicate the double-well shape. We increase the chemical potential in the normal region so that there are two transverse channels at the Fermi level (Fig.~\ref{fig2}(a)). We also add Dreseelhauss SOC with $\beta=0.2\alpha$. In this case, we plot the $E_{J\alpha}^{n}$ and $E_{J\beta}^{n}$ as a function of $M$ Fig~\ref{fig3}(a). Generally speaking, because both the time-reversal and inversion symmetries are broken due to the Zeeman effect and SOCs respectively, it allows finite $\sin\phi$ potential term which changes the double-well potential shape significantly (Fig.~\ref{fig3}(b)). Meanwhile, $\sin\phi$ potential in principle can lead the single Cooper pair tunneling even the $\cos\phi$ vanishes. At $n_g=0$, we plot the lowest two energy wave function distribution and the energy gap $E_{01}$ and $E_{02}$ in Fig.~\ref{fig3}(c) and Fig.~\ref{fig3}(d) respectively. Remarkably, the two lowest energy states remain almost degeneracy around $\cos\phi=0$ even in the presence of finite $\sin\phi$ while they are still the Cooper parity eigenstates. Although this result is calculated at $E^{(1)}_{j\beta}/E^{(2)}_{j\alpha} \approx 0.25$, which is obtained from the Fourier transform of the free energy, it remains valid even for larger $\sin\phi$ term. In Fig.~\ref{fig3}(e), we plot the probability of the lowest energy wave function distributing at even and odd number sites as a function of the $E^{(1)}_{J\beta}$. The lowest energy wave function only stay in the even number sites for $E^{(1)}_{j\beta}/E^{(2)}_{j\alpha}$ up to 0.4 (Fig.~\ref{fig3}(e)) and the coupling between even and odd states through $\sin\phi$ potential is zero (inset of Fig.~\ref{fig3}(e)). Meanwhile, the lowest eigenenergy is independent of the $\sin\phi$ potential. Therefore, the finite $\sin\phi$ term shows little effect on our qubit states at $n_g=0$. This is because the ground states are localized around either $\phi=0$ or $\phi=\pi$ for $E_c \ll E_{J\alpha}^{(2)}$ so that $\langle g| \sin(\hat{\phi}) | g \rangle \approx 0$. Therefore it affects the degenerate high excited states more than the low energy states. As a result, the qubit states with the potential in Fig.~\ref{fig3}(b) at $n_g=0$ can still be well described by Eq.~\eqref{wf}
which is robust against the imperfect perturbations and is essential for implementing 0-$\pi$ qubit.

{\it Coherence properties of 0-$\pi$ qubit - } To obtain the relaxation time of our 0-$\pi$ qubit, we first estimate the parameters in our system. For 2DEG we estimate $m*=0.026m_e,\Delta/h=45$GHz\cite{Larsen2020,Ren2019,Ke2019}, the electron density is about $n=10^{12}cm^{-2}$\cite{Ke2019,Liu2014} and the magnitude of magnetic field at 0-$\pi$ point satisfy the condition $g\mu_B B=\frac{1}{2}\cdot\frac{\pi}{2}\cdot\frac{\hbar v_f}{L}$\cite{Pientka2017,Ren2019,Fornieri2019}, gate voltage fluctuation is about $10^{-6} v$\cite{Casparis2018}, which shift the 0-$\pi$ point of magnetic field is smaller than distinguishability of magnetic field $10^{-5}T$, so we neglect the influence of gate voltage fluctuation. Since our realization of the double-well potential comes from the Zeeman effect induced spin splitting, the magnetic field is our main noise source. Besides, charge noise should also be considered. Noted that our 0-$\pi$ qubit relies on the sweet spot of both the two noises. We thus expand the Hamiltonian up to the second-order at the sweet spot.
\begin{equation}
H=H_0+\frac{\partial H}{\partial \lambda}\delta \lambda(t)+\frac{1}{2} \frac{\partial^2 H}{\partial \lambda^2}\delta \lambda^2(t)
\end{equation}
where $H_0$ is the Hamiltonian at $n_g=0$ with the well potential shown in Fig.~\ref{fig3}(b), and $\lambda$ represents the noise source and can be charge($n_g$) or magnetic field($B$). 

\begin{figure}[t]
\centerline{\includegraphics[width=1\columnwidth]{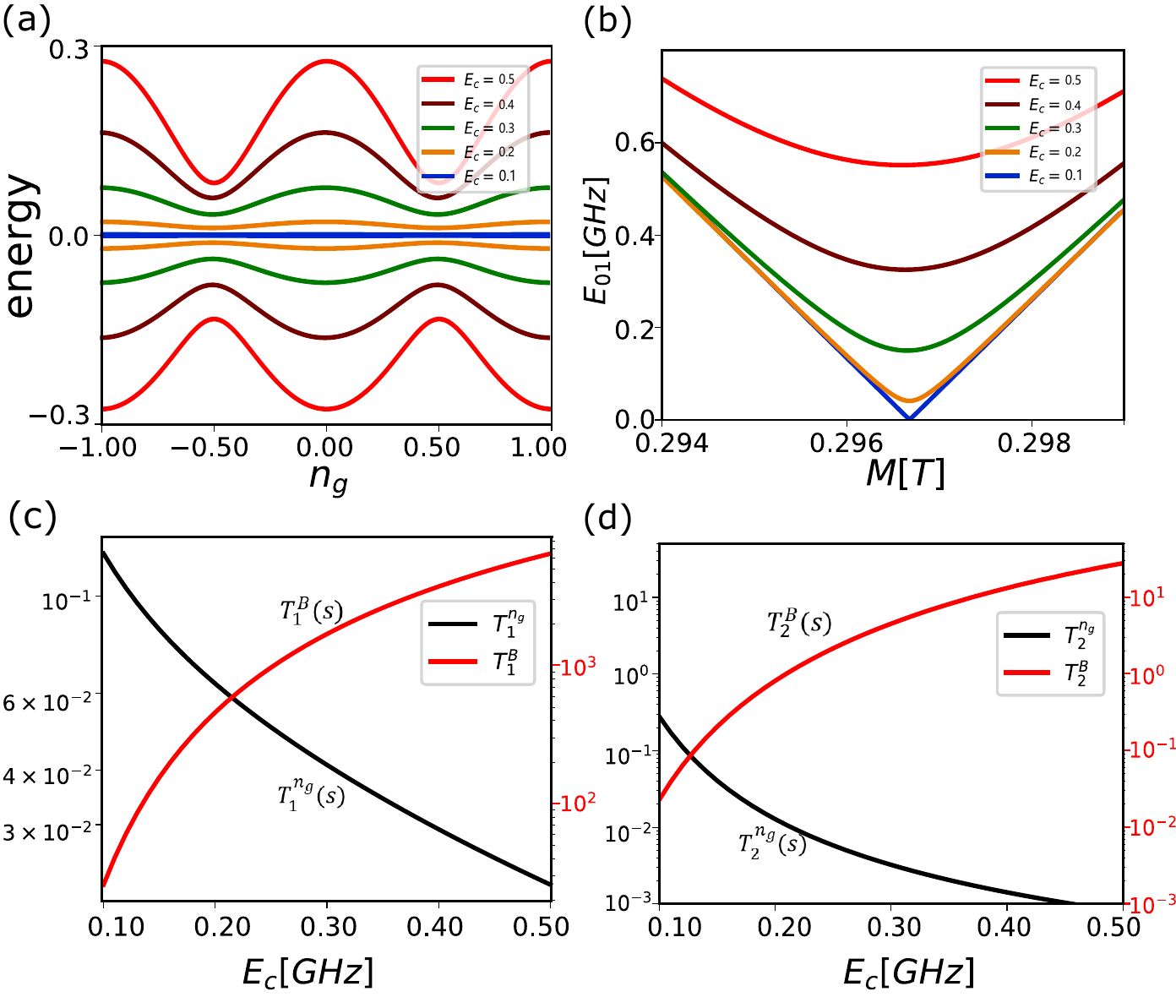}}
\caption{(a)Energy spectrum at 0-pi point with different $E_c$. (b)$E_{01}$ changes with magnetic field around 0-pi point with different $E_c$. (c)(d)$T_1,T_2$ changes with $E_c$ for charge noise and magnetic field noise, line in black correspond to left y-axis, line in red correspond to right y-axis.}
\label{fig4}
\end{figure}

According to the Fermi's golden rule \cite{Razeghi2019}, we can get $T_1$ from the inverse of the transition rate from initial state$|\psi_i\rangle$ to final state $|\psi_f\rangle$
\begin{equation}
\Gamma_{i\to f} = D_{\lambda \bot}^2 S_{\lambda}(\omega_{fi})
\end{equation}
with $D_{\lambda \bot} =\langle \psi_f| \frac{\partial H}{\partial \lambda}|\psi_i\rangle$ the transition amplitude and $S_{\lambda}(\omega)=\int_{-\infty}^{\infty}e^{-i\omega t}\langle \delta \lambda(t)\delta \lambda(0)\rangle$ is the noise power spectrum. For superconducting qubit it is typically approximate by 1/f spectrum, $S_{\lambda}(\omega)=2\pi\frac{A_\lambda}{|\omega|}(\omega_{ir}<\omega<\omega_{uv})$, where $A_{\lambda}$ is the noise amplitude for channel $\lambda$\cite{Ithier2005,Paladino2014}. Here we estimate that $\omega_{ir}/2\pi=1Hz$,$\omega_{uv}/2\pi=1GHz$ which is determinted by temperature(T$<$50mk), $A_{n_g}=10^{-8}e^2$\cite{Koch2007}, $A_B=10^{-17}T^2$\cite{Blais2007}. For charge noise, we have $\frac{\partial H}{\partial n_g}|_{n_g=0}=8E_c\hat{n}\delta n_g$. As the two 0-$\pi$ qubit states are the eigenstates of Cooper pair parity operator $
\hat{P}$, the qubit ordinary relaxation $\langle 0|\hat{n}|1\rangle$ vanishes. Therefore, the relaxation rate for charge noise is mainly from the depolarization rate \cite{Groszkowski2018,Krantz2019},
\begin{equation}
\Gamma_{1}^{n_g} = \Gamma_{0\to 2}^{n_g}+\Gamma_{0\to 3}^{n_g}+\Gamma_{1\to 2}^{n_g}+\Gamma_{1\to 3}^{n_g},
\end{equation}
which describes the transition speed from qubit states to higher levels.
The magnetic field noise is given by $\frac{\partial H}{\partial B}=\frac{\partial E^{(1)}_{j\alpha}}{\partial B}\cos\phi+\cdots$, which breaks the degeneracy of the double well and cause ordinary relaxation with the transition rate $\Gamma_1^{B}=|\langle 1|\frac{\partial H}{\partial B}|0\rangle|^2 S_B(\omega_{01})$.  the result shown in Fig~\ref{fig4}(c). It is mainly limited by charge noise, for $E_c=0.12$GHz, $T_1=91ms$.

The dephasing time $T_2$ is related to the decay of off-diagonal term of density matrix,
\begin{equation}
\rho_{01}=\exp(-i\frac{\partial \omega_{01}}{\partial \lambda}\int_0^t \delta \lambda(t)dt-i \frac{1}{2}\frac{\partial^2\omega_{01}}{\partial \lambda^2}\int_0^t\delta \lambda(t)^2dt),
\end{equation}
with $\omega_{01} = E_{01}/\hbar$. For calculation convenience, we use the notion $D_1 = \frac{\partial \omega_{01}}{\partial \lambda}, D_2 =\frac{\partial^2 \omega_{01}}{\partial \lambda^2}$. After standard calculation \cite{Groszkowski2018,Ithier2005}, we have
\begin{equation}
T_2 = [A_\lambda D_1^2ln(\frac{1}{\omega_{ir}t})+D_2^2A_\lambda^2ln^2(\frac{\omega_{uv}}{\omega_{ir}})+2D_2^2A_\lambda^2ln^2(\frac{1}{\omega_{ir}t})]^{-1/2}
\label{t2}
\end{equation}
In our system, $\omega_{01}$ is a sweet spot for charge($n_g$) and magnetic field(B) where the linear noise susceptibility vanishes. For second order term $\frac{\partial^2\omega_{01}}{\partial n_g^2}$, which can be suppressed by larger $E^{(2)}_{j\alpha}/E_c$ like Transmon\cite{Koch2007}, as shown in Fig~\ref{fig4}(a). However for magnetic field noise, the finite gap at 0-pi point is contributed by charge energy $E_c$, larger $E_c$ correspond to larger gap and smaller $\frac{\partial^2\omega_{01}}{\partial B^2}$ shown in Fig~\ref{fig4}(b). So increasing $E_c$, we can get larger $T_2^{n_g}$ but smaller $T_2^{B}$ shown in Fig~\ref{fig4}(d). We estimate the effective dephasing time is about $T_2=44ms$ at the cross point $E_c=0.12$GHz.


{\it Conclusions-} We propose to implement a 0-$\pi$ qubit in one SC/Sm/SC Josephson junction. Benefiting from the rich manipulability of the semiconductor, its internal spin degree of freedom naturally allow two 
Cooper pair interference paths, which have similar transport parameters such as transmission amplitude even under SOC and in-plane magnetic field. The qubit states are the eigenstates of Cooper pair parity operator and nearly degenerate which are robust against various deviations from the ideal model. The qubit relaxation time $T_1$ and coherent time $T_2$ can be dramatically increased. We expect that the multi-dimensional tunable SC/Sm/SC junction is a promising platform to realize parity-protected 0-$\pi$ qubit.

\section*{Acknowledge}
We would like to thank Jie Shen, Shun Wang, Cheng-Yu Yan and Dong E. Liu for fruitful discussions. X. Liu acknowledges the support of NSFC (Grant No.12074133), NSFC (Grant No.11674114) and National Key R\&D Program of China (Grant No. 2016YFA0401003). 

%

\end{document}